\documentclass[twocolumn,showpacs,preprintnumbers,amsmath,amssymb,aps]{revtex4}

\usepackage{graphicx}% Include figure files
\usepackage{dcolumn}% Align table columns on decimal point
\usepackage{bm}% bold math
\usepackage[colorlinks,linkcolor=blue,anchorcolor=blue,citecolor=blue]{hyperref}
\usepackage[caption=false,font=normalsize,labelfont=sf,textfont=sf]{subfig}
\usepackage{enumerate}
\usepackage{enumitem}
\usepackage{multirow}
\usepackage{amsmath}
\usepackage{booktabs}
\usepackage{amsmath,amsfonts}
\usepackage{bm}% bold math
\usepackage[ruled]{algorithm2e}
\begin{document}

%\preprint{APS/123-QED}

\title{Quantum Graph Convolutional Networks Based on Spectral Methods}% Force line breaks with \\

\author{Zi Ye$^{1,2}$}
\author{Kai  Yu$^1$}
%\thanks{Corresponding author. Email address:  @.edu.cn}
\author{Song Lin$^1$}%
\thanks{Corresponding author. Email address: lins95@fjnu.edu.cn}
%\altaffiliation{Corresponding author. Email address: lins95@gmail.com}
\affiliation{%
{$^1$College of Computer and Cyber Security, Fujian Normal University, Fuzhou 350117, China\\
$^2$School of Technology, Fuzhou Technology and Business University, Fuzhou 350117, China}
}%


\date{\today}% It is always \today, today,

\begin{abstract}
Graph Convolutional Networks (GCNs) are specialized neural networks for feature extraction from graph-structured data. In contrast to traditional convolutional networks, GCNs offer distinct advantages when processing irregular data, which is ubiquitous in real-world applications. This paper introduces an enhancement to GCNs based on spectral methods by integrating quantum computing techniques. Specifically, a quantum approach is employed to construct the Laplacian matrix, and phase estimation is used to extract the corresponding eigenvectors efficiently. Additionally, quantum parallelism is leveraged to accelerate the convolution operations, thereby improving the efficiency of feature extraction. The findings of this study demonstrate the feasibility of employing quantum computing principles and algorithms to optimize classical GCNs. Theoretical analysis further reveals that, compared to classical methods, the proposed quantum algorithm achieves exponential speedup concerning the number of nodes in the graph.
\end{abstract}

\maketitle

%\tableofcontents

\section{\label{sec:1}Introduction}
\par In the real world, many problems involve irregular data structures, such as social networks, logistics networks, protein interactions, and knowledge graphs. These data can be effectively represented using graph structures, which enable the encoding of complex geometric relationships and facilitate in-depth analysis through advanced machine learning techniques. Convolutional neural networks (CNNs) have demonstrated remarkable success in extracting meaningful patterns from large-scale datasets, particularly in image, sound, and video recognition. However, these tasks typically involve data with a regular grid-like structure, making traditional CNNs less suited for irregular data. To address this challenge, researchers have proposed graph convolutional neural networks (GCNs) tailored for graph-structured data \cite{1,2,3,4,5}. For instance, Bruna et al. introduced both spatial-domain and spectral-domain convolution methods for graph-structured data using spatial techniques and the graph Fourier transform \cite{1}. Defferrard et al. proposed an efficient filtering scheme that eliminates the need for explicit computation of Laplacian eigenvectors by leveraging Chebyshev polynomials in the spectral domain \cite{2}. Kipf et al. introduced a scalable graph neural network for semi-supervised learning, employing a first-order approximation to define the convolutional architecture \cite{3}.

\par In the era of big data, neural networks have become an essential tool in machine learning. However, as information technology continues to advance, the exponential growth of data volume imposes significant computational challenges on neural networks when processing high-dimensional data. The unique properties of quantum computing, particularly superposition and entanglement, offer significant computational advantages for handling high-dimensional problems, attracting considerable interest in the field of machine learning. However, since quantum computing is still in an exploratory stage, John et al. introduced the concept of noisy intermediate-scale quantum (NISQ) devices to describe the current generation of quantum hardware \cite{6,7}. To leverage the computational potential of quantum acceleration within the NISQ era, researchers have developed hybrid quantum-classical algorithms that integrate quantum and classical computing, making them compatible with NISQ devices \cite{8,9,10}. These approaches aim to enhance classical algorithms by incorporating quantum advantages. In this context, quantum neural networks (QNNs) have emerged as a promising paradigm, combining the strengths of both quantum computing and traditional neural networks, with the potential to significantly improve the computational efficiency of classical neural networks.

\par Meanwhile, quantum graph convolutional neural networks (QGCNs) \cite{11,12,13} have also undergone significant development. Zhang et al. introduced a quantum walk-based approach to enable weight sharing in QGCNs \cite{10}. Hu et al. proposed a fully quantum implementation of graph convolutional neural networks using variational quantum circuits, inspired by classical semi-supervised graph neural networks. Their results demonstrated that QGCNs exhibit potential advantages as the number of features increases \cite{11}. Bai et al. developed a quantum spatial graph convolutional neural network capable of directly learning classification functions for graphs of arbitrary size while preserving the lattice structure of input vertices. They further validated the model's effectiveness on a benchmark graph classification dataset \cite{12}. Qu et al. employed variational methods to construct a quantum convolutional neural network \cite{13}; however, the variational circuits in this algorithm do not explicitly leverage the structural properties of graph data. In this work, we integrate the graph Laplacian matrix with a quantum-classical hybrid approach to design a quantum graph convolutional neural network (QGCN) for spectral methods, building upon the existing spectral-based graph convolutional neural network framework \cite{1}.

\par The remainder of this paper is organized as follows. The classical graph convolutional neural network is reviewed in Sec. \ref{sec:2}. In Sec. \ref{sec:3}, we propose the framework for QGCN. In Sec. \ref{sec:4}, we analyze the time complexity of QGCN. In Sec. \ref{sec:5}, we give the conclusion of our work.

\section{\label{sec:2}Review of classical graph convolutional neural network} %Overview of research foundations

\par Classical Graph Convolutional Neural Networks (CGCNs) are a type of deep learning model specifically designed for processing graph-structured data. CGCNs perform convolutions on graphs, allowing them to capture relationships between nodes based on the graph topology. CGCNs are closely related to the graph Laplacian matrix, particularly in spectral-based methods. The graph Laplacian matrix is a fundamental mathematical representation of a graph that captures its structural properties. The eigenvalues and eigenvectors of the Laplacian matrix provide insights into the the structure of the graph, connectivity, and clustering properties.

\par A CGCN based on the spectral method applies the principles of spectral graph theory to process graph-structured data. Instead of directly aggregating neighboring node features like spatial methods, spectral GCNs define graph convolutions using the graph Fourier transform, which relies on the eigenvalues and eigenvectors of the graph Laplacian matrix.In this approach, graph signals (node features) are transformed into the spectral domain, where convolution operations are performed as filtering processes.

\par For For a graph convolutional neural network (GCN) [1] based on the spectral method, consider an undirected graph $G=(V, E)$ withn nodes. The vertex set $V$ is defined as $V={{\textbf{x}_i^0|\textbf{x}_i^0\in R^{f_0}}}_{i=0}^{n-1}$ for $i=0,\cdots,n-1,$ where $f_0$ is the number of node features. Additionally, the weight matrix $W\in R^{n\times n}$ associated with the graph $G$ , which is defined as:

%Eq1
     \begin{equation}
     %\begin{split}
        { w_{ij}=w_{ji}=\left\{
           \begin{aligned}
            \geq0,\ \ \ \ \ i\neq j \\
           =0,\ \ \ \ i=j
          \end{aligned}
           \right.},
     %\end{split}
     \label{eq:1}
     \end{equation}

where $w_{ij}$ represents an entry in the weight matrix W. The degree matrix $D\in R^{n\times n}$ of the graph $G$ is a diagonal array, defined as $D=diag\left(d_{ii}\right)$ , where $d_{ii}=\sum_{j=0}^{n-1}w_{ij}$. The Laplace matrices $L\in R^{n\times n}$ of the graph $G$ is given by $L=D-W$ .

\par Consider a neural network with a total of $S$ convolutional layers, where $f_s$ denotes the number of features in the $\textbf{s}$-th layer. The convolution process from layer $s$ to $s+1(s=1\ldots S)$ is defined as follows. First, the Laplacian matrix $L$ is computed using the degree matrix and the adjacency matrix. The matrix $V$ consists of the eigenvectors of $L$, arranged in columns according to the following equation:

%Eq2
     \begin{equation}
     %\begin{split}
        \textbf{x}_j^s=h\left(V\sum_{i=1}^{f_{s-1}}F_{s,i,j}V^T\textbf{x}_i^{s-1}\right)\left(j=1\ldots\ f_s\right)
     %\end{split}
     \label{eq:2}
     \end{equation}

where $\textbf{x}_{k,i}$ represents the vector of the $i$-th feature for each node in the $s$-th layer, and $x_{s+1,j}$ represents the vector of the $j$-th feature for each node in the $s+1$-th layer. $F_{s,i,j}$ is the convolution kernel that operates on the $i$-th feature of the $s$-th layer to produce the $j$-th feature, and each $F_{s,i,j}$ is a diagonal matrix, defined as $F_{s,i,j}=diag[\theta_1,\ \ \theta_2,\ \ \cdots,\ \ \theta_d]$, where $h$ denotes the activation function. The column vectors of $V$ are the eigenvectors of the Laplacian matrix, arranged in descending order of their corresponding eigenvalues. In practice, only the eigenvectors associated with the d largest eigenvalues are retained. In this case, $V$ in Equation \ref{eq2} can be replaced by $V_d$ which consists of the first d columns of $V$. Consequently, the input to the next graph convolution layer is given by:

%Eq3
     \begin{equation}
     %\begin{split}
        \textbf{x}_j^s\ =h\left(V_d\sum_{i=1}^{f_{s-1}}F_{k,i,j}{V_d}^T\textbf{x}_i^{s-1}\right)\left(j=1\cdots f_s\right)
     %\end{split}
     \label{eq:3}
     \end{equation}

%figure 1
    %\begin{figure}
       % \centering
        %\includegraphics[width=0.46\textwidth]{FL}% Here is how to import EPS art
        %\caption{Schematic diagram of the federated learning based on gradient descent. $\mathbf{w}^{j}(n)$ is denoted the $j$th element of the parameter vector $ \mathbf{w}(n)$. % %\bm{G}^{j} \left( \mathbf{w}(n) \right)$ is the $j$th component of the global gradient. $\beta_{k}$ is the aggregation weight. $\bm{g}^{j}_{k}\left( \mathbf{w}(n) \right)$ is represented as the $j$th element of local gradient of the client ${\rm Bob}_{k}$. $\alpha$ is a learning rate.}
   % \label{fig:1}
   % \end{figure}
%%%

%
%%%%%%%%%%%%%%%Sec3%%%%%%%%%%%%%%%%%%%%%
\section{\label{sec:3}Quantum graph neural network based on spectral approach}

\par Consider the following unitary operations used to efficiently prepare the matrix $X_{M\times N}\in R^{{M\times N}}$. Let $x_{mn}$ denote the element in the $n$-th column and $m$-th row of $X$, while $\textbf{x}_{m,\bullet}$ represents the $m$-th row vector and $\textbf{x}_{\bullet,n}$ represents the $n$-th column vector of $X$. The unitary transformations are defined as follows:
%Eq4
     \begin{equation}
     %\begin{split}
        U_1:\left.|m\right\rangle\left.|0\right\rangle\rightarrow\frac{1}{\sqrt M}\sum_{m}{\left.|m\right\rangle\frac{1}{\sqrt{{||\textbf{x}_{m,\bullet}||}_2}}\sum_{n}{x_{mn}\left.|n\right\rangle}},
     %\end{split}
     \label{eq:4}
     \end{equation}

%Eq5
     \begin{equation}
     %\begin{split}
        U_2:\left.|n\right\rangle\left.|0\right\rangle\rightarrow\frac{1}{\sqrt N}\sum_{n}\left.|n\right\rangle\frac{1}{\sqrt{{||\textbf{x}_{\bullet,n}||}_2}}\sum_{m}\left.x_{mn}|m\right\rangle.
     %\end{split}
     \label{eq:5}
     \end{equation}
		
When the matrix elements are stored using a binary tree structure in quantum random access memory (QRAM) \cite{10,11}, these operations can be implemented in $O(poly\log_2N)$ time.

%%%

\subsection{\label{sec:3.1}Extracting the eigenvectors of the graph Laplace matrix}

\par Since the Laplacian matrix of a graph is an ergodic matrix, its corresponding density operator can be derived using quantum Hamiltonian simulation algorithms, provided that the Laplacian matrix is known. Currently, researchers have proposed various quantum Hamiltonian simulation algorithms for ergodic matrices, including those designed for sparse Hamiltonian \cite{17,18,19} and non-sparse Hamiltonian \cite{20}.

\par If the Laplacian matrix is unknown, its density operator can still be constructed using quantum algorithms designed to generate the Laplacian matrix based on node information \cite{21,22,23}. Some of these methods require knowledge of the weight and degree matrices. However, in most cases, the graph Laplacian matrix can be prepared solely from vertex information. Notably, when the adjacency matrix is constructed using the Gaussian similarity function, the Laplacian matrix becomes dense. Reference [23] introduces methods for constructing the unitary operator $e^{iLt}$ of the Laplacian matrix using only node information.

\par The quantum phase estimation of $e^{iLt}$ yields the decomposition:
%Eq6
     \begin{equation}
         L^\prime=\sum_{j=1}^{n-1}{\lambda_j\left|\left.v_j\right\rangle\right.\left\langle\left.v_j\right|\right.\otimes\left|\left.\lambda_j\right\rangle\right.\left\langle\left.\lambda_j\right|\right.}
     \label{eq:6}
     \end{equation}

\par By performing $O(d)$ measurements on $\left|\left.\lambda_j\right\rangle\right.$, the $d$ largest eigenvalues $\lambda_1,\ \ \lambda_2,\ \ \cdots,\lambda_d$ can be obtained with high probability. Furthermore, the quantum states of the eigenvectors $\left|\left.v_j\right\rangle\right.$, corresponding to these eigenvalues, can also be extracted.
\subsection{\label{sec:3.2}Extracting node feature information}

\par (2.1) Prepare a quantum register in state $\left.|0\right\rangle_1^{\otimes\log_2f_s}\otimes\left.|0\right\rangle_2^{\otimes\log_2d}\otimes\left.|0\right\rangle_3^{\otimes\log_2n}\otimes\left.|0\right\rangle_4$, where \(\left|0\right\rangle^a = \underbrace{\left|0\right\rangle\cdots\left|0\right\rangle}_{a}\).
Operation \(H^{\otimes\log_2 f_s}\) and operation \(H^{\otimes\log_2 d}\) are performed on registers 1 and 2, respectively.
Operation \(H\) is performed on register 4, then the system becomes

%Eq7
     \begin{equation}
        \frac{1}{\sqrt{f_s}}\sum_{j=0}^{f_s-1}{\left.|j\right\rangle_1\frac{1}{\sqrt d}\sum_{k=0}^{d-1}\left.|k\right\rangle_2}\left.|0\right\rangle_3\frac{1}{\sqrt2}{(\left.|0\right\rangle}_4+\left.|1\right\rangle_4).
     \label{eq:7}
     \end{equation}
	
\par (2.2) Using the operation $U_2$, the node feature matrix $X_{n\times f_k}={(x_1,\ \ x_2,\ \ \cdots,\ \ x_n)}^T$ is prepared under the control of register 1 and register 4. Simultaneously, the feature vector matrix $V_d$, which has dimensions $n\times d$, is prepared under the control of register 2 and 4. At this stage, the system state is given by:
	
%Eq8
     \begin{equation}
        \frac{1}{\sqrt{f_s}}\sum_{j=0}^{f_s-1}{\left.|j\right\rangle_1\frac{1}{\sqrt d}\sum_{k=0}^{d-1}\left.|k\right\rangle_2(\frac{1}{\sqrt2}\left.|x_{\bullet,j}\right\rangle_3}\left.|0\right\rangle_4+\frac{1}{\sqrt2}\left.|\textbf{v}_{\bullet,k}\right\rangle_3\left.|1\right\rangle_4),	
     \label{eq8}
     \end{equation}
	
where $\left.|\textbf{x}_{\bullet,j}\right\rangle_3=\frac{1}{\sqrt{{||\textbf{x}_{\bullet,j}||}_2}}\sum_{i=0}^{n-1}{x_{ij}\left.|i\right\rangle_3} ,\left.|\textbf{v}_{\bullet,k}\right\rangle_4=\frac{1}{\sqrt{{||\textbf{v}_{\bullet,k}||}_2}}\sum_{l=0}^{n-1}{v_{lk}\left.|l\right\rangle_3}$ .

\par The operation $H$ is performed on register 3, the system becomes:

%Eq9
     \begin{equation}
        \frac{1}{\sqrt{f_s}}\sum_{j=0}^{f_s-1}{\left.|j\right\rangle_1\frac{1}{\sqrt d}\sum_{k=0}^{d-1}\left.|k\right\rangle_2}\left|\left.\psi_{jk}\right\rangle\right._{34},
     \label{eq9}
     \end{equation}
		
which $\left|\left.\psi_{jk}\right\rangle\right._{34}=\frac{1}{2}\left(\left.|\textbf{x}_{\bullet,j}\right\rangle_3+\left.|\textbf{v}_{\bullet,k}\right\rangle_3\right)\left.|0\right\rangle_3+\frac{1}{2}\left(\left.|\textbf{x}_{\bullet,j}\right\rangle_3-\left.|\textbf{v}_{\bullet,k}\right\rangle_3\right)\left.|1\right\rangle_4 $.

\par The operations to obtain the quantum state $\left.|\psi_n\right\rangle$ can be denoted as $\mathcal{A}_{jk}$, i.e.$\mathcal{A}_{jk}\left.|0\cdots0\right\rangle=\left.|\psi_{jk}\right\rangle$. The state$\left.|\psi_{kj}\right\rangle$ can be re-described as:
	
%Eq10
     \begin{equation}
        \left.|\psi_{jk}\right\rangle=cos\theta_n\left.|\psi_{jk}^0\right\rangle+sin\theta_n\left.|\psi_{jk}^1\right\rangle,
     \label{eq10}
     \end{equation}
where

%Eq11
     \begin{equation}
       \left.|\psi_{jk}^0\right\rangle=\frac{1}{2cos\theta_{jk}}\left(\left.|\textbf{x}_{\bullet,j}\right\rangle_3+\left.|\textbf{v}_{\bullet,k}\right\rangle_3\right)\left.|0\right\rangle_4,
     \label{eq11}
     \end{equation}

%Eq12
     \begin{equation}
      \left.|\psi_{jk}^1\right\rangle=\frac{1}{2sin\theta_{jk}}\left(\left.|\textbf{x}_{\bullet,j}\right\rangle_3-\left.|\textbf{v}_{\bullet,k}\right\rangle_3\right)\left.|1\right\rangle_4,
     \label{eq12}
     \end{equation}

And it can be verified
%Eq13
     \begin{equation}
     cos{(\theta}_{jk})=\frac{1}{\sqrt2}\sqrt{1+\left\langle\textbf{x}_{\bullet,j}\middle|\textbf{v}_{\bullet,k}\right\rangle}.
     \label{eq13}
     \end{equation}

\par Define the Grover operator $Q_{jk}$,  incorporating the specified state-flip operation within the superposition state and the diffusion operation as proposed by the Grover algorithm.
%Eq14
     \begin{equation}
      \begin{aligned}
        Q_{jk}&=\mathcal{A}_{jk}S_0\mathcal{A}_{jk}^\dag S_0^\prime\\
        &=-\left(I - 2\left|\psi_{jk}\right\rangle\left\langle\psi_{jk}\right|\right)\left(I - {2I}^{\otimes2\log_2n}\otimes\left|1\right\rangle\left\langle1\right|\right)
    \end{aligned}
     \label{eq14}
     \end{equation}

where $S_0=-(I-{2\left.|0\right\rangle\left\langle0|\right.}^{\otimes2n+1})$ .
\par There exist two eigenvalues $\lambda_\pm=e^{\pm2i\theta_{jk}}(i=\sqrt{-1})$ and two corresponding eigenvectors $\left.|\psi_{jk}^\pm\right\rangle=\frac{1}{\sqrt2}(\left.|\psi_{jk}^0\right\rangle\mp\left.|\psi_{jk}^1\right\rangle)$ for this operator $Q_{jk}$, at which point $\left.|\psi_n\right\rangle$ can be represented on the set of eigenvectors:

%Eq15
     \begin{equation}
     \left.|\psi_{jk}\right\rangle=\frac{1}{\sqrt2}\left(e^{i\theta_{jk}}\left.|\psi_{jk}^+\right\rangle+e^{-i\theta_n}\left.|\psi_{jk}^-\right\rangle\right).
     \label{eq15}
     \end{equation}

\par (2.3) Phase estimation. Add the register 5 in state $\left.|0\right\rangle^{\otimes q}$, where $q=O(log1/\epsilon)$, $\epsilon$ is the error. Perform the operation $H^{\otimes q}$ on register 5 to get:
%Eq16
     \begin{equation}
    \frac{1}{\sqrt{f_s}}\sum_{j=0}^{f_s-1}{\left.|j\right\rangle_1\frac{1}{\sqrt d}\sum_{k=0}^{d-1}\left.|k\right\rangle_2}\left|\left.\psi\right\rangle\right._{34}\sum_{s=0}^{2^q-1}\left.|s\right\rangle_5.
     \label{eq16}
     \end{equation}

\par Using registers 1, 2, and 5 as control registers, the controlled $Q_{jk}^{2^s}$ operation is applied to registers 2, 3, and 4. This is followed by an inverse Fourier transform on register 5, ultimately yielding an approximation of the eigenvalue of $Q_{jk}$ in register 5. At this stage, the system state is:
	
%Eq17
     \begin{equation}
     \begin{aligned}
        \left|\mathrm{\Psi}_4\right\rangle&=\frac{1}{\sqrt{2f_s}}\sum_{j = 0}^{f_s - 1}{\left|j\right\rangle_1\frac{1}{\sqrt{d}}\sum_{k = 0}^{d - 1}\left|k\right\rangle_2}{e^{i\theta_n}\left|\psi_{jk}^+\right\rangle\left|\widetilde{\theta}_{jk}\right\rangle}_{345}\\
        &+\frac{1}{\sqrt{2f_s}}\sum_{j = 0}^{f_s - 1}{\left|j\right\rangle_1\frac{1}{\sqrt{d}}\sum_{k = 0}^{d - 1}\left|k\right\rangle_2}{e^{-i\theta_n}\left|\psi_{jk}^-\right\rangle\left|-\widetilde{\theta}_{jk}\right\rangle}_{345}
    \end{aligned}
     \label{eq17}
     \end{equation}

which is ${\widetilde{\theta}}_{jk}\in Z_{2^q}$ and satisfies $|\theta_{jk}-{\widetilde{\theta}}_{jk}\pi/2^q|\le\epsilon$ .

\par (2.4) Information extraction. Since $cos({\widetilde{\theta}}_{jk})=cos(-{\widetilde{\theta}}_{jk})$ , performing cosine-gate[24], the transformation $\left.|{\widetilde{\theta}}_{jk}\right\rangle\rightarrow\left.|cos({\widetilde{\theta}}_{jk})\right\rangle$ can be realized to obtain the quantum state:

%Eq18
     \begin{equation}
      \begin{aligned}
        \left.|\mathrm{\Psi}_5\right\rangle=\frac{1}{\sqrt{{2f}_s}}\sum_{j=0}^{f_s-1}{\left.|j\right\rangle_1\frac{1}{\sqrt d}\sum_{k=0}^{d-1}\left.|k\right\rangle_2}\\
        \left(e^{i\theta_n}\left.|\psi_{jk}^+\right\rangle+ e^{-i\theta_n}\left.|\psi_{jk}^-\right\rangle\right)_{34}\left.|cos({\widetilde{\theta}}_{jk})\right\rangle_5
    \end{aligned}
     \label{eq18}
     \end{equation}
The inverse of the phase estimation operation is performed on register 4, the remaining system state is:

%Eq19
     \begin{equation}
   \frac{1}{\sqrt{f_s}}\sum_{j=0}^{f_s-1}{\left.|j\right\rangle_1\frac{1}{\sqrt d}\sum_{k=0}^{d-1}\left.|k\right\rangle_2}\left.|cos({\widetilde{\theta}}_{jk})\right\rangle_5.
     \label{eq19}
     \end{equation}

Adding register 6 as the auxiliary register, the same state is obtained using CNOT gate under the control of register 5, and according to the quantum multiplier,${\cos{\left({\widetilde{\theta}}_{jk}\right)}}^2$ is obtained in register 5, at which point the system state is:
%Eq20
     \begin{equation}
   \frac{1}{\sqrt{f_s}}\sum_{j=0}^{f_s-1}{\left.|j\right\rangle_1\frac{1}{\sqrt d}\sum_{k=0}^{d-1}\left.|k\right\rangle_2}\left.|{\cos{\left({\widetilde{\theta}}_{jk}\right)}}^2\right\rangle_5\left.|\cos{\left({\widetilde{\theta}}_{jk}\right)})\right\rangle_6.
     \label{eq20}
     \end{equation}	

Due to $\left\langle x_{\bullet,j}\middle| v_{\bullet,k}\right\rangle=2{\cos{\left({\widetilde{\theta}}_{jk}\right)}}^2-1$, the state in register 6 is $\left.|2\right\rangle_6$ utilizing quantum Paui-X gate. Using quantum multiplier, $2{\cos{\left({\widetilde{\theta}}_{jk}\right)}}^2$ can be obtained in register 5. By the same way, the value in register 6 is set to 1, using quantum multiplier, one can get:

%Eq21
     \begin{equation}
  \frac{1}{\sqrt{f_s}}\sum_{j=0}^{f_s-1}{\left.|j\right\rangle_1\frac{1}{\sqrt d}\sum_{k=0}^{d-1}\left.|k\right\rangle_2}\left.\left|\left\langle\textbf{x}_{\bullet,j}\middle|\textbf{v}_{\bullet,k}\right\rangle\right.\right\rangle_5\left.|1\right\rangle_6.	
     \label{eq21}
     \end{equation}

%%%%%%%%%%%%%%%%%%

\subsection{\label{sec:3.3} Variational Convolution}

\par (3.1) $F_{s,i,j}$ is a convolution kernel represented as a diagonal matrix, where the diagonal elements are ${\theta_1,\ \ \theta_2,\ \ \cdots,\ \ \theta_d}$ and all off-diagonal elements are zero. To incorporate these parameters, register 7 is added, are loaded under the control of register 2. At this stage, the system state is given by:
	
%Eq22
     \begin{equation}
  \frac{1}{\sqrt{f_s}}\sum_{j=0}^{f_s-1}{\left.|j\right\rangle_1\frac{1}{\sqrt d}\sum_{k=0}^{d-1}\left.|k\right\rangle_2}\left.\left|\left\langle\textbf{x}_{\bullet,j}\middle|\textbf{v}_{\bullet,k}\right\rangle\right.\right\rangle_5\left.|\theta_k\right\rangle_6.	
     \label{eq22}
     \end{equation}	

Following the quantum multiplication operation [21], multiplication is performed under the control of register 2, leading to the updated system state: 	

%Eq23
     \begin{equation}
  \frac{1}{\sqrt{f_s}}\sum_{j=0}^{f_s-1}{\left.|j\right\rangle_1\frac{1}{\sqrt d}\sum_{k=0}^{d-1}\left.|k\right\rangle_2}\left.\left|\left\langle\textbf{x}_{\bullet,j}\middle|\textbf{v}_{\bullet,k}\right\rangle\right.\right\rangle_5\left.|\theta_k\right\rangle_6\left.\left|\theta_k\left\langle\textbf{x}_{\bullet,j}\middle|\textbf{v}_{\bullet,k}\right\rangle\right.\right\rangle_7.
     \label{eq23}
     \end{equation}	

\par (3.2) This step is computed using a quantum adder \cite{21}, controlled by registers 1 and 2, with the result stored in quantum register 8:

%Eq24
     \begin{equation}
     \begin{aligned}
  \frac{1}{\sqrt{f_s}}\sum_{j=0}^{f_s-1}{\left.|j\right\rangle_1\frac{1}{\sqrt d}\sum_{k=0}^{d-1}\left.|k\right\rangle_2}
  \left.\left|\left\langle\textbf{x}_{\bullet,j}\middle|\textbf{v}_{\bullet,k}\right\rangle\right.\right\rangle_5\left.|\theta_k\right\rangle_6\left.\left|\theta_k\left\langle\textbf{x}_{\bullet,j}\middle|\textbf{v}_{\bullet,k}\right\rangle\right.\right\rangle_7\\
  \left.\left|\sum_{j=0}^{f_s-1}{\theta_k\left\langle\textbf{x}_{\bullet,j}\middle|\textbf{v}_{\bullet,k}\right\rangle}\right.\right\rangle_8.
  \end{aligned}
  \label{eq24}
     \end{equation}	

\par (3.3) Uncompute the step of multiplication and discard the register 6 and register 7,

%Eq25
     \begin{equation}
 \frac{1}{\sqrt{f_s}}\sum_{j=0}^{f_s-1}{\left.|j\right\rangle_1\frac{1}{\sqrt{d-1}}\sum_{k=0}^{d-1}{\left.|k\right\rangle_2\left.\left|\left\langle\textbf{x}_{\bullet,j}\middle|\textbf{v}_{\bullet,k}\right\rangle\right.\right\rangle_5}}\left.\left|\sum_{j=0}^{f_s-1}{\theta_k\left\langle\textbf{x}_{\bullet,j}\middle|\textbf{v}_{\bullet,k}\right\rangle}\right.\right\rangle_8.	   \label{eq25}
     \end{equation}	

The convolutional kernel parameters are encoded using CNOT gates, forming a key component of a variational circuit. This setup enables the tuning of the convolutional kernel in a hybrid quantum-classical manner by adjusting the applied CNOT gates.

%%%%%%%%%%%%%%%%%%%%%%%%%%%%%%%%%%%%%sec.4%%%%%%%%%%%%%%%%%%%%%%%%%%%%%%%%%%%%%%%%%%%%%%%
\subsection{\label{sec:3.4} Getting the lower convolutional layer inputs}

\par (4.1) Add the register in state $\left.|0\right\rangle_9$ to perform a controlled rotation, then, the system state is:

%Eq26
    {\footnotesize
\begin{equation}
    \begin{aligned}
        &\frac{1}{\sqrt{f_s}}\sum_{j = 0}^{f_s - 1} \left|j\right\rangle_1
        \left( \frac{1}{\sqrt{d}} \sum_{k = 0}^{d - 1} \left|k\right\rangle_2
        \left| \left\langle \bm{x}_{\bullet,j} \middle| \bm{v}_{\bullet,k} \right\rangle \right\rangle_5
        \left| \sum_{j = 0}^{f_s - 1} \theta_k \left\langle \bm{x}_{\bullet,j} \middle| \bm{v}_{\bullet,k} \right\rangle \right\rangle_8
        \eta_k \left|0\right\rangle_9 \right. \\
        &+ \left. \frac{1}{\sqrt{d}} \sum_{k = 0}^{d - 1} \left|k\right\rangle_2
        \left| \left\langle \bm{x}_{\bullet,j} \middle| \bm{v}_{\bullet,k} \right\rangle \right\rangle_5
        \left| \sum_{j = 0}^{f_s - 1} \theta_k \left\langle \bm{x}_{\bullet,j} \middle| \bm{v}_{\bullet,k} \right\rangle \right\rangle_8
        \sqrt{1 - \eta_k} \left|1\right\rangle_9 \right)
    \end{aligned}
    \label{eq26}
\end{equation}
}

where $\eta_k=\frac{1}{C}\sum_{j=0}^{f_k-1}{\theta_k\left\langle x_{\bullet,j}\middle| v_{\bullet,k}\right\rangle}$. Uncompute the previous steps,

%Eq27
     \begin{equation}
    \frac{1}{\sqrt d}\sum_{k=0}^{d-1}{\eta_k\left.|k\right\rangle_2\left.|0\right\rangle_9}+\sqrt{1-\eta_k}\left.|k\right\rangle_2\left.|1\right\rangle_9.
   \label{eq27}
     \end{equation}	

\par (4.2) The eigenvector matrix $V_d$ is prepared by expanding the registers using the operation $U_1$, The resulting system state is:

%Eq28
     \begin{equation}
     \begin{aligned}
    \frac{1}{\sqrt d}\sum_{p=0}^{d-1}\left.|p\right\rangle_A\frac{1}{\sqrt{{||\textbf{v}_{p,\bullet}||}_2}}\sum_{q=0}^{n-1}{v_{pq}\left.|q\right\rangle_B}\\
    \frac{1}{\sqrt d}\sum_{k=0}^{d-1}{(\eta_k\left.|k\right\rangle_2\left.|0\right\rangle_9}+\sqrt{1-\eta_k}\left.|k\right\rangle_2\left.|1\right\rangle_9).
    \end{aligned}
   \label{eq28}
     \end{equation}	

\par (4.3) Using auxiliary quantum bits for the exchange test, the system state can be expressed as:

%Eq29
     \begin{equation}
   \frac{1}{\sqrt d}\sum_{p=0}^{d-1}\left.|p\right\rangle_A\left.|\textbf{v}_{p,\bullet}\right\rangle_B\left.|f\right\rangle_2\left.|0\right\rangle_9+\left.|{\zeta_1}^\bot\right\rangle,	
   \label{eq29}
     \end{equation}

where $\left.|f\right\rangle_2=\frac{1}{\sqrt d}\sum_{k=0}^{d-1}{\eta_k\left.|k\right\rangle_2}$ , and $\left.|\textbf{v}_{p,\bullet}\right\rangle_B=\frac{1}{\sqrt{{||\textbf{v}_{p,\bullet}||}_2}}\sum_{q=0}^{n-1}{v_{pq}\left.|q\right\rangle_B}$ .
Next, a Hadamard ($H$) operation is applied to register 9. A Swap operation is then performed on registers 2 and B, controlled by register 9, implementing the transformation $\left.|a\right\rangle\left.|b\right\rangle\rightarrow\left.|b\right\rangle\left.|a\right\rangle$. Finally, another Hadamard operation is applied to register 9, resulting in the system state:
	
%Eq30
     \begin{equation}
   \frac{1}{\sqrt{d-1}}\sum_{p=0}^{d-1}\left.|p\right\rangle_A\left.|\phi_p\right\rangle+\left.|{\zeta_2}^\bot\right\rangle,
   \label{eq30}
     \end{equation}	

where

%Eq31
     \begin{equation}
     \begin{aligned}
  \left.|\phi_p\right\rangle=\frac{1}{2}\left(\left.|v_{p,\bullet}\right\rangle_B\left.|f\right\rangle_2+\left.|f\right\rangle_B\left.|\textbf{v}_{p,\bullet}\right\rangle_2\right)\left.|0\right\rangle_9\\
  +\frac{1}{2}\left(\left.|\textbf{v}_{p,\bullet}\right\rangle_B\left.|f\right\rangle_2-\left.|f\right\rangle_B\left.|\textbf{v}_{p,\bullet}\right\rangle_2\right)\left.|1\right\rangle_9	
   \end{aligned}
  \label{eq31}
     \end{equation}	

Denoting the sequence of operations that produce the quantum state $\left.|\psi_n\right\rangle as \mathcal{A}_n$, we can express:$\mathcal{A}_p\left.|0\cdots0\right\rangle=\left.|\psi_p\right\rangle.$ The state $\left.|\psi_n\right\rangle$ can be rewritten as:
%Eq32
     \begin{equation}
	\left.|\phi_p\right\rangle=cos\alpha_p\left.|\psi_{p0}\right\rangle+sin\theta_n\left.|\psi_{p1}\right\rangle,
    \label{eq32}
     \end{equation}	

where

%Eq33
     \begin{equation}
 \left.|\phi_{p0}\right\rangle=\frac{1}{2cos\alpha_p}\left(\left.|f\right\rangle_2\left.|\textbf{v}_{p,\bullet}\right\rangle_B+\left.|\textbf{v}_{p,\bullet}\right\rangle_2\left.|f\right\rangle_B\right)\left.|0\right\rangle_9,	 \label{eq33}
     \end{equation}	

%Eq34
     \begin{equation}
 \left.|\phi_{p1}\right\rangle=\frac{1}{2sin\alpha_p}\left(\left.|f\right\rangle_2\left.|\textbf{v}_{p,\bullet}\right\rangle_B-\left.|\textbf{v}_{p,\bullet}\right\rangle_2\left.|f\right\rangle_B\right)\left.|1\right\rangle_9,	 \label{eq34}
     \end{equation}	

It follows that:

%Eq35
     \begin{equation}
   cos{(\alpha}_p)=\frac{1}{\sqrt2}\sqrt{1+\left(f\bullet\textbf{v}_{p,\bullet}\right)^2}.
    \label{eq35}
     \end{equation}	

By combining the specified state-flip operation in the superposition state with the superposition-state diffusion operation introduced by Grover's algorithm, the defined Grover operator $G_p$ is expressed as:

%Eq36
     \begin{equation}
   G_p=\mathcal{A}_pS_0\mathcal{A}_p^\dag S_0^\prime=-\left(I-2\left.|\phi_p\right\rangle\left\langle\phi_p|\right.\right)\left(I-{2I}^{\otimes2d}\otimes\left.|1\right\rangle\left\langle1|\right.\right),
    \label{eq36}
     \end{equation}	

Here, $S_0=-(I-{2\left.|0\right\rangle\left\langle0|\right.}^{\otimes2d+1})$ .

\par There exist two eigenvalues $\lambda_\pm=e^{\pm2i\alpha_p}(i=\sqrt{-1})$ and two corresponding eigenvectors $\left.|\phi_p^\pm\right\rangle=\frac{1}{\sqrt2}(\left.|\phi_{n0}\right\rangle\mp\left.|\phi_{n1}\right\rangle)$ for this operator $G_p$, Thus, the state $\ \left.|\phi_n\right\rangle$ can be re-expressed in terms of these eigenvectors as:
	
%Eq37
     \begin{equation}
   \left.|\phi_p\right\rangle=\frac{1}{\sqrt2}\left(e^{i\alpha_p}\left.|\phi_p^+\right\rangle+e^{-i\alpha_p}\left.|\phi_p^-\right\rangle\right).
    \label{eq37}
     \end{equation}

\par (4.4) Introduce an auxiliary register C initialized to the state $\left.|0\right\rangle^{\otimes q}$ where $q=O(\log_2(\frac{1}{\epsilon}))$, with $\epsilon$ representing the error. The phase estimation algorithm is applied by performing a Hadamard operation $H^{\otimes q}$ on register 5, yielding:
%Eq38
  \begin{equation}
     \frac{1}{\sqrt d}\sum_{p=0}^{d-1}\left.|p\right\rangle_A\left.|\phi_p\right\rangle\sum_{u=0}^{2^q-1}\left.|u\right\rangle_C+\left.|\zeta^\bot\right\rangle.
    \label{eq38}
  \end{equation}

Using registers $A$ and $C$ as control registers, a controlled $Q_p^{2^u}$ is applied to registers $B$, 2, and 9. The inverse quantum Fourier transform is then performed on register $C$, yielding an approximation of the eigenvalue of $Q_p$ stored in register $C$. The resulting system state is:
	
%Eq39
     \begin{equation}
	\frac{1}{\sqrt d}\sum_{p=0}^{d-1}\left.|p\right\rangle_A\left(e^{i\alpha_p}\left.|\phi_p^+\right\rangle\left.|{\widetilde{\alpha}}_p\right\rangle+e^{-i\alpha_p}\left.|\phi_p^-\right\rangle\left.|-{\widetilde{\alpha}}_p\right\rangle\right)+\left.|\zeta^\bot\right\rangle, \label{eq39}
     \end{equation}	

Where ${\widetilde{\alpha}}_p\in Z_{2^q} satisfies |\alpha_p-{\widetilde{\alpha}}_p\pi/2^q|\le\epsilon$.
\par (4.5) Since $cos({\widetilde{\alpha}}_p)=cos(-{\widetilde{\alpha}}_p)$, a cosine-gate \cite{25} is applied to transform $\left.|{\widetilde{\alpha}}_p\right\rangle\rightarrow\left.|cos({\widetilde{\alpha}}_p)\right\rangle$, resulting in the quantum state:

%Eq40
     \begin{equation}
     \begin{aligned}
     \left.|\mathrm{\Psi}_5\right\rangle_{12345}=\frac{1}{\sqrt{2N}}\sum_{p=0}^{n-1}\left.|p\right\rangle_A\\
     \left(e^{i\alpha_p}\left.|\phi_p^+\right\rangle+e^{-i\alpha_p}\left.|\phi_p^-\right\rangle\right)\left.|cos({\widetilde{\alpha}}_p)\right\rangle_C+\left.|\zeta^\bot\right\rangle
	 \end{aligned}
     \label{eq40}
     \end{equation}	

Next, measuring register 9 in the $\left.|0\right\rangle$ state collapses the system into:

%Eq41
     \begin{equation}
\frac{1}{\sqrt{2N}}\sum_{p=0}^{n-1}\left.|p\right\rangle_A\left(e^{i\alpha_p}\left.|\phi_p^+\right\rangle+e^{-i\alpha_p}\left.|\phi_p^-\right\rangle\right)\left.|\cos{\left({\widetilde{\alpha}}_p\right)})\right\rangle_C.		\label{eq41}
     \end{equation}	

The next layer of input information is stored as quantum states in register $C$. By iterating this process, a multi-layer quantum convolution operation can be realized in a graph convolutional neural network.

\section{\label{sec:4}Analysis}
\par According to the Laplace matrix preparation algorithm [23], the time complexity of constructing the Laplace matrix L and extracting its eigenvalues is given by $O(d\sqrt npoly\log_2nf_k)$. In the convolutional computation process, the primary source of time complexity arises from phase estimation, which is predominantly influenced by the construction of the Grover operator. The time complexity of constructing the Grover operator is $O(poly\log_2(n))$. The overall time complexity of phase estimation depends on both the complexity of Grover operator construction and the estimation error $\epsilon$. When the error is $\epsilon$, the time complexity of phase estimation is $O(poly\log_2(n)/\epsilon)$
\par Furthermore, based on quantum quadrature \cite{25}, the time complexity of quantum multiplication and quantum addition is $O(poly\log_2(n))$. As a result, the time complexity of a single convolution operation is $O(d\sqrt npoly\log_2(nf_k)/\epsilon)$. The algorithm proposed in this paper achieves an exponential speedup in the number of nodes compared to the time complexity of classical graph convolutional neural networks.

\section{\label{sec:5}Conclusions}
\par This paper presents a quantum graph convolutional neural network algorithm based on the spectral approach. The algorithm leverages quantum parallelism in feature extraction to reduce computational complexity. It extracts the feature vectors of the graph Laplace matrix using phase estimation and employs the exchange test to perform convolutional computations between these feature vectors and the node features, generating the input for the next convolutional layer. Theoretical analysis demonstrates that, compared to classical algorithms, the proposed quantum algorithm achieves an exponential speedup in terms of the number of graph data nodes. However, optimizing the algorithm for numerical simulation remains an open research direction in graph convolutional neural networks.

%\bibliography{QGCN}

\end{document}